\documentclass{ws-procs9x6}

\usepackage[english]{babel}

\begin{document}

\title{The NA62 Liquid Krypton\\
Electromagnetic Calorimeter\\
Level 0 Trigger}

\author{A. Fucci, G. Paoluzzi, A. Salamon$^*$, G. Salina}

\address{INFN Sezione di Roma Tor Vergata\\
Via della Ricerca Scientifica, 1 - 00133 Roma Italia\\
$^*$E-mail: andrea.salamon@roma2.infn.it}

\author{E. Santovetti, F. M. Scarf\`{\i}}

\address{Universit\`a degli Studi di Roma Tor Vergata - Dip. di Fisica\\
Via della Ricerca Scientifica, 1 - 00133 Roma Italia}

\author{V. Bonaiuto, F. Sargeni}

\address{Universit\`a degli Studi di Roma Tor Vergata - Dip. di Ingegneria Elettronica\\
Via del Politecnico, 1 - 00133 Roma Italia}

\begin{abstract}
The NA62 experiment at CERN SPS aims to measure the Branching Ratio of the very rare kaon decay $K^+\to\pi^+\nu\bar{\nu}$ collecting $O(100)$ events with a 10\% background to make a stringent test of the Standard Model. One of the main backgrounds to the proposed measurement is represented by the $K^+ \to \pi^+\pi^0$ decay. To suppress this background an efficient photo veto system is foreseen. In the 1-10 mrad angular region the NA48 high performance liquid krypton electromagnetic calorimeter is used. The design, implementation and current status of the Liquid Krypton Electromagnetic Calorimeter Level 0 Trigger are presented.
\end{abstract}

\bodymatter

\section{The NA62 experiment at CERN SPS}
The NA62 {experiment \cite{cit1,cit2}} aims at measuring the very rare kaon decay $K^+\to\pi^+\nu\bar{\nu}$ collecting $O(100)$ events with a 10\% background in two years of data taking.

$K^+\to\pi^+\nu\bar{\nu}$ is an exceptionally clean decay from the theoretical point of view with a Standard Model branching ratio prediction of BR($K^+\to\pi^+\nu\bar{\nu}$) = $(8.5 \pm 0.7) \times 10^{-11}$. A precise measurement of BR($K^+\to\pi^+\nu\bar{\nu}$) will offer the opportunities of testing the Standard Model and deepening the knowledge of the CKM matrix.

\begin{figure}
\vspace{-0.3cm}
\begin{center}
\includegraphics[totalheight=0.22\textheight,viewport=75 178 1450 700,clip]{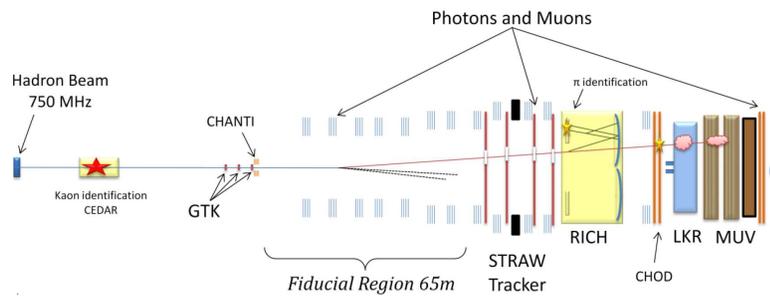}
\end{center}
\vspace{-0.3cm}
\caption{Schematic drawing of the NA62 detector at CERN SPS.}
\label{fig1}
\vspace{-0.3cm}
\end{figure}

The NA62 {detector \cite{cit3}}, see Figure \ref{fig1}, currently being installed at the SPS North Area High Intensity Facility is composed of: a differential Cerenkov {counter \cite{cit4}} (CEDAR), a beam tracker (GTK) and charged particle detector (CHANTI), a straw chambers magnetic spectrometer, a photon veto system composed of different detectors in the various angular decay regions, a {RICH \cite{cit5}}, a charged particle hodoscope (CHOD) and a muon detector (MUV).

\section{Trigger and Data Acquisition system}
In order to extract few interesting decays from a very intense flux a complex and performing three level trigger and data acquisition system was {designed \cite{cit6}}.

The Level 0 trigger algorithm is based on few sub-detectors and is performed by dedicated custom hardware modules, with a maximum output rate of 1 MHz and a maximum latency of 1 ms.

Level 1 and Level 2 software triggers are executed on dedicated PCs. The maximum Level 2 output rate is of the order of 15 kHz.

\section{The Liquid Krypton electromagnetic calorimeter}
In order to suppress the background from $K^+ \to \pi^+\pi^0$ decay an efficient photon veto system is foreseen. In the 1-10 mrad angular region the NA48 electromagnetic calorimeter is {used \cite{cit7}}.

This calorimeter is a quasi-homogenous ionization device using liquid krypton as active medium and characterized by excellent time and energy resolution.

The Liquid Krypton calorimeter will be readout by the new Calorimeter REAdout {Modules \cite{cit8}} (CREAMs) which will provide 40 MHz 14 bit sampling for all 13248 calorimeter readout channels, data buffering, optional zero suppression and programmable trigger sums for the Level 0 electromagnetic calorimeter trigger processor.

\section{The Liquid Krypton Level 0 trigger}
\begin{figure}
\vspace{-0.3cm}
\begin{center}
\includegraphics[totalheight=0.19\textheight,viewport=10 180 715 390,clip]{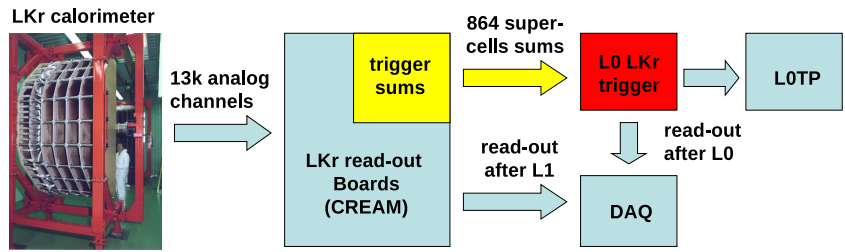}
\end{center}
\vspace{-0.3cm}
\caption{Block diagram of the Liquid Krypton Level 0 trigger inside the Trigger and Data Acquisition system.}
\label{fig2}
\vspace{-0.3cm}
\end{figure}

The Level 0 Liquid Krypton electromagnetic calorimeter trigger, see Figure \ref{fig2},  identifies electromagnetic clusters in the calorimeter and prepares a time-ordered list of reconstructed clusters together with the arrival time, position, and energy measurements of each cluster. Information on reconstructed clusters is used by the Level 0 Trigger Processor to veto decays with more than one cluster in the Liquid Krypton calorimeter.

The trigger processor also provides a coarse-grained readout of the Liquid Krypton calorimeter that can be used in software triggers and off-line as a cross-check for the CREAM high-granularity readout. 

\subsection{Trigger algorithm}
Trigger algorithm is based on energy deposits in tiles of 16 calorimeter cells which are available from the main readout boards.

\begin{figure}
\vspace{-0.3cm}
\begin{center}
\includegraphics[totalheight=0.23\textheight,viewport=00 75 720 455,clip]{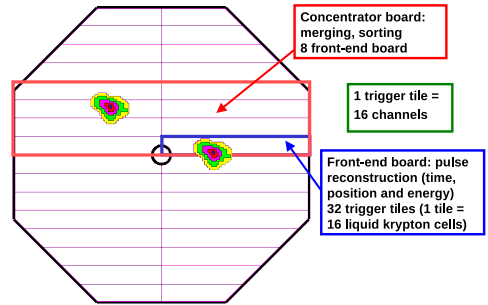}
\end{center}
\vspace{-0.3cm}
\caption{Liquid Krypton electromagnetic calorimeter trigger segmentation.}
\label{fig3}
\vspace{-0.3cm}
\end{figure}

Electromagnetic cluster search is executed in two steps with two one-dimensional (1D) algorithms, see Figure \ref{fig3}.

From the trigger point of view the calorimeter is divided in slices parallel to the horizontal axis. In the first step peaks in space and time are searched independently in each slice with a 1D algorithm. In the second step different peaks which are close in time and space are merged and assigned to the same electromagnetic cluster.

\subsection{Trigger processor  implementation}
The main parameters driving the design of the processor are the high expected instantaneous hit rate (30 MHz), the required single cluster time resolution (1.5 ns) and a maximum allowed latency of 100 $\mu$s.

\begin{figure}
\vspace{-0.3cm}
\begin{center}
\includegraphics[totalheight=0.24\textheight,viewport=0 135 720 435,clip]{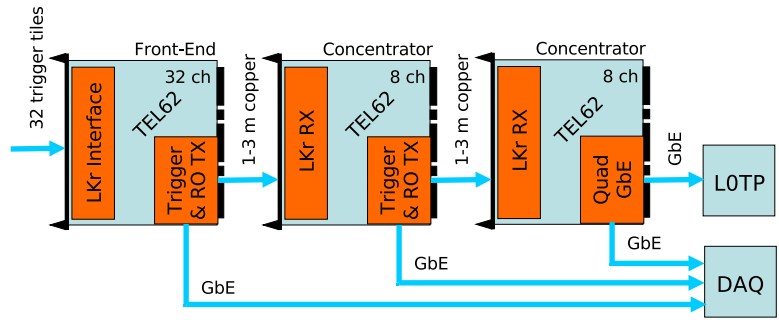}
\end{center}
\vspace{-0.3cm}
\caption{Liquid Krypton trigger processor block diagram. 28 Front-End boards and 8 Concentrator boards are foreseen in the system.}
\label{fig4}
\vspace{-0.3cm}
\end{figure}

The processor is a three-layer parallel system, composed of Front-End and Concentrator boards, both based on the 9U TEL62 {cards \cite{cit9}} equipped with custom dedicated mezzanines, see Figure \ref{fig4}.

The Liquid Krypton Level 0 trigger continuously receives from the Liquid Krypton readout modules 864 trigger sums\footnote{Trigger sums are transmitted over shielded copper twisted pairs.} each one corresponding to a tile of 16 calorimeter cells.

Each {\bf Front-End board} receives 32 trigger sums and performs peak search in space and computes time, position and energy for each detected peak. In order to extract timing information at the ns level a parabolic interpolation in time around sample maximum and a digital constant fraction discrimination are performed after the peak search algorithms. Information on reconstructed peaks is transferred from the Front-End boards to the Concentrator boards on low-latency high-bandwidth dedicated trigger links. Raw data received by the readout modules are also stored in latency memories, to be readout upon request. 28 Front-End boards equipped with 84 custom mezzanines are foreseen in the whole system.

The {\bf Concentrator board} receives trigger data from up to 8 FE boards and combines peaks detected by different front-end boards into a single cluster. Overlap between neighboring Concentrators is foreseen to guarantee that each cluster will be fully contained in at least one Concentrator board with proper logic to avoid double counting. Reconstructed clusters are also stored in latency memories, to be readout upon request. 8 Concentrator boards equipped with 24 custom mezzanines are foreseen in the whole system.

\section*{Conclusions}
A fast parallel processor for cluster reconstruction and counting in the Liquid Krypton electromagnetic calorimeter of the NA62 experiment has been designed.

In total, the system will be composed of 36 TEL62 boards, 108 mezzanine cards and 215 high-performance FPGAs. The whole system will fit in three 9U crates.

The peak reconstruction algorithm was implemented on FPGA and was proven to fulfill the NA62 timing and rate requirements. Most of the boards and mezzanines have been designed and are currently being tested and assembled together for data taking.

\renewcommand{\bibname}{References}

\end{document}